\documentclass[prd,amssymb,twocolumn,floatfix]{revtex4}
\usepackage{graphicx}

\begin{document}
\preprint{CERN-PH-TH/2007-158}
\title{The zero age main sequence of WIMP burners}

\author{Malcolm Fairbairn}
\affiliation{PH-TH, CERN, Geneva, Switzerland \& King's College London, WC2R 2LS, UK}
\email{malc@cern.ch}
\author{Pat Scott and Joakim Edsj\"o}
\affiliation{Cosmology, Particle Astrophysics and String Theory, Physics, Stockholm University \&\\
High Energy Astrophysics and Cosmology Centre (HEAC),AlbaNova Univeristy Centre, SE-106 91 Stockholm, Sweden}
\email{pat@physto.se, edsjo@physto.se}

\begin{abstract}
We modify a stellar structure code to estimate the effect upon the main sequence of the accretion of weakly interacting dark matter onto stars and its subsequent annihilation.  The effect upon the stars depends upon whether the energy generation rate from dark matter annihilation is large enough to shut off the nuclear burning in the star.  Main sequence WIMP burners look much like protostars moving on the Hayashi track, although they are in principle completely stable.  We make some brief comments about where such stars could be found, how they might be observed and more detailed simulations which are currently in progress.  Finally we comment on whether or not it is possible to link the paradoxically young OB stars found at the galactic centre with WIMP burners.
\end{abstract}

\maketitle

There is growing evidence that for each gram of baryonic matter in the universe, there are around five grams of dark matter which does not couple to the electromagnetic force \cite{wmap}.  One of the more convincing candidates for this dark matter are weakly interacting massive particles (WIMPs).  Because WIMPs have masses and couplings close to the weak scale, they naturally give rise to a relic abundance of dark matter close to that observed after freeze out in the early universe.

Because WIMPs couple weakly to standard model particles, there is a small but non-zero WIMP-nucleon cross section.  The tightest constraints for the spin-independent cross section come from the XENON experiment \cite{xenon}.  The WIMP-nucleon interaction means that some WIMPs are gravitationally captured by stars, a process which has been well studied \cite{accrete, gould}.   Usually the focus of such investigations is the possibility of dark matter annihilating into high energy neutrinos, which then escape the star to be potentially detected by neutrino experiments like IceCube.  If the accretion of dark matter were large enough however, one might expect that the stars themselves could change.  

The first way WIMPs can affect a star is by annihilating into standard model particles in its core, providing another energy source in addition to standard nuclear burning.  We refer to stars where the energy produced by WIMP annihilations is greater than or comparable to that from nuclear burning as `WIMP burners'.  Recent work has focused upon white dwarfs and neutron stars \cite{moskwai,malcbert}.  Here we instead focus on main sequence WIMP burners.  Previous work in this direction was carried out by Salati \& Silk \cite{salatisilk} some years ago, in the context of cosmion dark matter.  These authors used an $n=3$ polytropic approximation to estimate the influence of WIMP annihilation on the main sequence.  We instead employ full numerical solutions to the hydrostatic equations of stellar structure and present results from a simple code, as well as supporting preliminary results from a more advanced code.  We also update the discussion in the context of `modern WIMPs' rather than cosmions, as the nuclear scattering cross-sections and the WIMP mass are more constrained than in the past.

The second way WIMPs can influence stellar structure is by providing an additional mechanism of heat transport in the core.  This could reduce the local temperature gradient \cite{transport}, potentially inhibiting convection and enhancing the pulsation of horizontal branch stars \cite{bouquet1,dearborn}.  At least for the Sun however, the values currently favoured for the WIMP self-annihilation cross section \cite{wmap} and the upper limit on the WIMP-nucleon cross section \cite{xenon} indicate that this effect is not significant \cite{bottino}. In this work, we neglect heat transport by WIMPs, but will in a later work include this effect in more detailed, time-dependent simulations \cite{future}. For the current work, we have checked that this omission does not affect our results significantly.

Experiments give relatively weak constraints on the spin-dependent WIMP-nucleon cross section, so it may contribute far more to WIMP capture than the spin-independent one.  For simplicity, we assume that the spin-dependent cross section dominates and is $\sigma_\mathrm{SD}=10^{-38}$\,cm$^2$. For a Sun-like star with mass $1\,M_\odot$, a circular velocity of 220\,km\,s$^{-1}$, a WIMP velocity dispersion of $\bar{v}=270$\,km\,s$^{-1}$ and a WIMP mass of 100\,GeV, the capture rate (as calculated with the full capture expressions \cite{gould} in DarkSUSY \cite{darksusy}) is then
\begin{equation}
\Gamma_0 = 2.90 \cdot 10^{24} \mbox{~s$^{-1}$}.
\end{equation}
We assume a hydrogen mass fraction of 75\%, uniformly distributed in the star.  This is not a good approximation for the Sun, but reasonable for newly born stars. The capture rate on an arbitrary star is then approximately
\begin{eqnarray}
\lefteqn{\Gamma_\mathrm{c} = \left(\frac{M_*}{M_\odot}\right)
\left(\frac{v_{\rm esc}}{618\,\mathrm{km\,s}^{-1}} \right)^2
\left(\frac{270\,\mathrm{km\,s}^{-1}}{\bar{v}}\right) } \nonumber \\
& & 
\mbox{} \times \left( \frac{\rho_\mathrm{w}}{0.3\,\mathrm{GeV\,cm}^{-3}} \right)
\left(\frac{100\,\mathrm{GeV}}{m_\mathrm{w}} \right)
\left( \frac{\langle \phi \rangle}{\langle \phi \rangle_\odot} \right) \Gamma_0,
\label{eq:caprate}
\end{eqnarray}
where $M_*$ is the star's mass and $v_{\rm esc}$ its surface escape velocity, $\rho_\mathrm{w}$ is the ambient WIMP density and $m_\mathrm{w}$ is the WIMP mass. $\langle \phi \rangle = \langle v_{\rm esc}^2(r)/v_{\rm esc}^2 \rangle$ is the average potential at the height of the scattering nuclei (hydrogen in this case, for which $\langle \phi \rangle_\odot= 3.16$).  For simplicity, we assume $\langle \phi \rangle / \langle \phi \rangle_\odot=1$.  Even if this approximation is not perfect, Eq. (\ref{eq:caprate}) is only necessary for converting between the ambient WIMP density $\rho_\mathrm{w}$ and the capture rate $\Gamma_{\rm c}$, so it is easy to rescale if the reader wishes.  We also assume $\bar{v}=270$\,km\,s$^{-1}$, though for stars in very different environments to our own this must be adjusted.  Throughout this paper, we use a WIMP mass of $100$\,GeV although to a first approximation the energy accreted by a star immersed in some fixed density of dark matter is only a function of the cross section, not the mass of the WIMP.

Once dark matter is captured by the star, it will form an approximately thermal (Gaussian) internal distribution, of characteristic radius \cite{bouquet2,bouquet1,bottino}
\begin{equation}
r_{\mathrm{w}} = \left[\frac{3kT_\mathrm{c}}{2 \pi G \rho_\mathrm{c} m_{\mathrm{w}}}\right]^{1/2},
\end{equation}
where $\rho_\mathrm{c}$ and $T_\mathrm{c}$ are the central density and temperature.  Having been concentrated in the centre of the star, the dark matter annihilates with itself, so the equation for the evolution of the number of WIMPs in the star over time is
\begin{equation}
\frac{dN}{dt}=\Gamma_\mathrm{c}-2\Gamma_\mathrm{a}.
\label{evol}
\end{equation}
When $\Gamma_\mathrm{c}=2\Gamma_\mathrm{a}$, the capture and annihilation rates are in equilibrium and any dark matter which accretes onto the star is instantly converted into additional luminosity.  Here we assume that all of the annihilation products interact either electromagnetically or strongly, so that they have short mean free paths in the star and the energy thermalises quickly.  In reality, some fraction of the energy will be lost to neutrinos, but this is a small effect (typically on the order of 10\% or less).  The timescale for the steady state to be reached will be much less than the typical evolutionary timescale of a main sequence star, so we assume that this equilibrium has been achieved.

We studied the effect of this energy input upon stars by altering the FORTRAN code ZAMS \cite{ZAMS}, which looks for stable stellar solutions assuming a constant chemical composition and a simple equation of state.  The code was modified by adding the energy generation due to WIMP annihilation to the nuclear energy generation rate.

Self gravitating systems have negative specific heat, so as dark matter injects energy into the core of the star, the temperature goes down.  Since the nuclear reaction rates depend exponentially on temperature, they are significantly reduced for even a small drop in temperature.  As the ambient density of dark matter is increased and the capture rate goes up, the injection of energy due to WIMP annihilations eventually reduces the temperature enough to shut off nuclear burning.  WIMP annihilation is then the primary source of the star's luminosity.

Since WIMP annihilation occurs in a more centralised region than nuclear burning, the temperature gradient is much steeper in the core of the star than it would otherwise be, and the core becomes convective.  Outside the core, less energy is generated per unit volume than if nuclear burning were proceeding normally, so the temperature gradient is less.  The actual temperature is lower than in a normal star, but for small dark matter accretion rates it remains high enough to prevent any major increase in opacity, ensuring that energy transport in the region above the core remains radiative.  The energy from the core is easily transmitted through this radiative zone to the surface layers of the star.  The overall radius of the star remains approximately constant while the temperature decreases, so the luminosity also decreases.

\begin{figure}
\hspace{2.5cm}
\begin{center}
\includegraphics[width=\columnwidth]{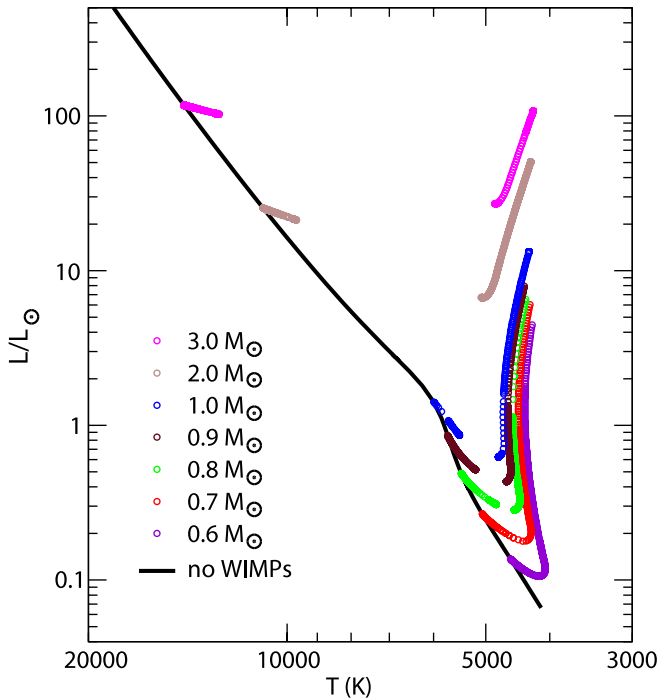} 
\caption{\it The zero age main sequence of WIMP burners.  The black solid line is the normal zero age main sequence for stars with solar metallicity.  The points moving off the main sequence correspond to solutions where stars contain increasing amounts of dark matter in their cores.  The gaps in the lines are addressed in the text.}
\label{hr}
\end{center}
\end{figure}

As the WIMP accretion rate is raised, the star continues to cool until H$^{-}$ ions are able to survive at increasingly large depths below the surface.  If the star did not already have an appreciable surface convection zone, one develops.  At high enough WIMP capture rates, the surface convection zone merges with the inner zone and the star becomes completely convective.  The addition of more WIMPs and hence central luminosity beyond this point requires the star to grow in order to transport the additional energy to the surface.  The two distinct stages of this `evolution' can clearly be seen in Figure \ref{hr}.
This figure shows a black solid line corresponding to the zero age main sequence, along with `evolutionary tracks' plotted as the ambient density of WIMPs is increased.  These tracks are strongly reminiscent of the Hayashi tracks which proto-stars travel along on the final stages of their evolution towards the main sequence.  The difference is that such stars are shrinking in size on the Kelvin-Helmholtz timescale as they radiate away gravitational potential energy.  In WIMP burners, there is constant energy generation in the core (provided the WIMP capture rate remains constant) and the stars can in principle remain at that position in the HR diagram for an arbitrarily long time.

The analysis of Salati \& Silk \cite{salatisilk} led to qualitatively similar conclusions about the backing up of the main sequence along the Hayashi track with increasing WIMP density.  The shapes of the tracks are different in our more detailed analysis, and thus the actual positions of the resulting WIMP burners in the HR diagram differ.  The effects require higher ambient WIMP densities also, as we have considered modern WIMPs rather than cosmions (i.e.~with lower nuclear scattering cross-sections).

\begin{figure}
\hspace{2.5cm}
\begin{center}
\includegraphics[width=\columnwidth]{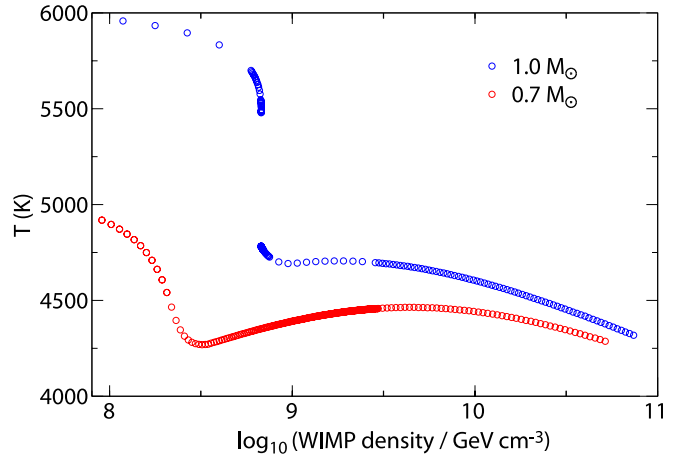} 
\caption{\it The temperature of main sequence stars of mass 1\,$M_\odot$ (upper curve) and 0.7\,$M_\odot$ as a function of WIMP density.  A spin dependent WIMP-nucleon cross section of $\sigma=10^{-38}$\,cm$^2$ is assumed.}
\label{temp}
\end{center}
\end{figure}

In Figure \ref{hr}, tracks for stars of mass 0.8\,$M_\odot$ are not complete, and show increasingly large gaps for larger masses.  Using the modified ZAMS code, we could not find solutions corresponding to all temperatures between the main sequence stars and the cooler Hayashi-like WIMP burners.  In Figure \ref{temp} we plot temperature as a function of ambient WIMP density.  The discontinuity occurs where the temperature of the star drops rapidly, shortly before the star becomes fully convective.  

It was not initially clear whether this lack of solutions was a real effect or an artefact of the code.  We have put some effort into understanding whether some of the physical simplifications in the code (e.g.~surface boundary conditions, a simplified equation of state and thus too rigid a criterion for the onset of surface convection, or the treatment of convective heat transport itself) are responsible for the lack of solutions.  We could find no evidence to support this idea.  Nonetheless, the gaps do not appear to have a physical basis.  They seem to be a numerical artefact arising from the `shooting' technique used by the ZAMS code to solve the boundary-value problem of stellar structure.  In the region of the gaps, small changes in the stellar radius cause large changes in the internal temperature.  The surface defined by the discrepancy between the inwardly and outwardly integrated partial solutions employed in this technique then becomes a highly non-trivial function of the model parameters.  Finding the global minimum of this function is then nearly impossible without an initial guess extremely close to the true solution, and the algorithm fails to converge.

\begin{figure}
\hspace{2.5cm}
\begin{center}
\includegraphics[width=\columnwidth]{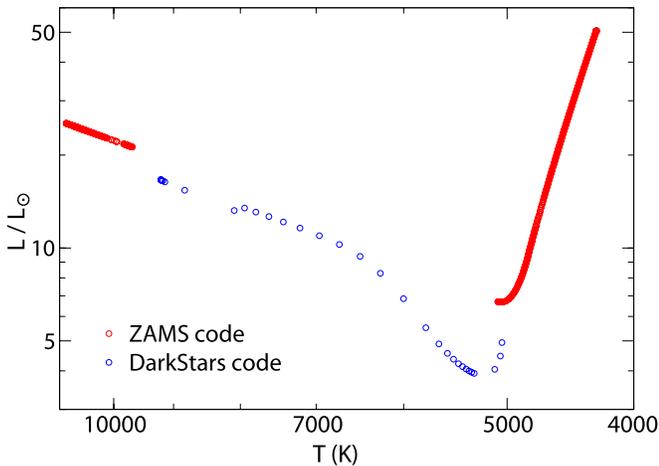} 
\caption{\it Diagram for $M_* = 2M_\odot$ showing that solutions can be found corresponding to all temperatures when the more sophisticated DarkStars code is used.}
\label{pat}
\end{center}
\end{figure}

To check that solutions do exist in this region, we investigated the gaps using a preliminary version of our next-generation WIMP burner code DarkStars.  This code has been created from generalised versions of capture routines in DarkSUSY \cite{darksusy} which use the full capture rate expressions in \cite{gould}, and the stellar evolution package EZ \cite{ez} derived from Eggleton's \textsc{stars} code \cite{stars}.  The code is time-dependent, uses relaxation rather than shooting and includes WIMP energy transport, more detailed treatments of the WIMP distribution, capture rates, equation of state, nuclear reaction rates and opacities.  DarkStars and results obtained with it will be described in full in an upcoming publication \cite{future}.  The results of the more detailed code (Figure \ref{pat}), whilst consistent with the ZAMS code close to the main sequence and in the fully convective regions on the right-hand side of the HR diagram, show that intermediate solutions do exist and that the gaps are not physical. This cross-check also shows that we can trust the solutions of the ZAMS code, but shouldn't take the actual gaps in the curves seriously.

Figure \ref{temp} shows that for a spin-dependent WIMP-nucleon cross section of $\sigma=10^{-38}$\,cm$^2$, stars only start to change their behaviour when immersed in a dark matter density of around $10^8$ or $10^9$\,GeV\,cm$^{-3}$.  This is much larger than the 0.3\,GeV\,cm$^{-3}$ which is thought to be the approximate density of dark matter in the solar system.  Most modern simulations of galactic dark matter halos do suggest that the density of dark matter should be much higher in the centre of the galaxy.  This density should be made more pronounced by the phenomena of adiabatic contraction and dark matter spikes created around black holes, though tempered by self-annihilation of dark matter and gravitational heating due to the motion of stars. It is suggested \cite{merritt} that densities larger than $10^8$\,GeV\,cm$^{-3}$ can be found at radii closer than $10^{-2}$\,pc from the galactic centre.  Observational challenges associated with constructing an HR diagram from stars in this crowded field are severe though, as the very centre of the galaxy is shrouded in dust.

Finally, we comment upon the `paradox of youth' implied by what appear to be hot, young stars close to the central black holes of M31 and our own galaxy \cite{ghez, demarque}.  In our galaxy, these stars look like $\sim$15\,$M_\odot$ main sequence stars with luminosities of the order 1,000\,$L_\odot$.  Such a massive main sequence star should only be around 10 Myr old, so these objects would appear to have formed very recently.  However, tidal forces close to the central black hole are thought to be too large to be compatible with the collapse of a gas cloud to begin star formation.  It is not easy to make normal UBV measurements of stellar temperatures at the galactic centre due to extinction, but spectral lines found in the atmosphere of such stars appear to be consistent with a temperature of 30,000 K \cite{ghez}.  If this is the case, then it would be difficult to imagine that these are normal stars which have been converted into fully convective and highly luminous WIMP burners from the accretion of dark matter, as such stars would have a temperature of only a few thousand degrees.

A possible explanation for the `paradox of youth' is a combination of collisional stripping of red giant envelopes and merger events \cite{demarque}, owing to the high stellar densities near the centres of the two galaxies.  If this is the case, then the anomalously hot stars in fact derive from an older population.  Such a population should therefore also contain less luminous, unperturbed stars, including lower mass stars still on the main sequence.  Provided that this explanation is correct, and if main sequence WIMP burners exist at all, then such lower mass stars should be examples of them.  With upcoming observations of the galactic centre expected to probe objects as faint as 1\,$L_\odot$, a very real possibility exists for the detection of WIMP burners in the near future.

\section*{Acknowledgments}
We thank Gianfranco Bertone, Torsten Bringmann and Georges Meynet for useful discussions. J.E.\ thanks the Swedish Research Council (VR) for support.

\end{document}